# Optimal lofted B-spline surface interpolation based on serial closed contours

Shutao Tang [*]

Department of Mechanical Engineering, Northwestern Polytechnical University, Xi'an, China

## Abstract

Modern shape design and capture techniques often lead to the geometric data presented in the form of serial rows of data points. In general, the number of data points varies from row to row. Lofted or skinned B-spline surface interpolation is a technique that generates a B-spline surface that passes through these data points precisely. The traditional process often causes a large increase in the number of control points of the resulting B-spline surface. Much of the work to date in mitigating the effects of this increase has been restricted to open section-curves. The lofting of sequential closed contours using the interpolation technique has not been addressed in the existing literature. In this paper, we present two novel conjectures relating to closed B-spline curve interpolation. We derive the equivalent closed B-spline interpolation condition of the well-established Schoenberg-Whitney condition for open B-spline interpolation, a condition that the parameter values and the domain knots should satisfy to guarantee the system matrix is always invertible or full-rank. We then apply the interpolation condition to the problem of lofted B-spline surface interpolation to serial closed contours. The correctness of these conjectures is validated via numerical results and several practical experiments.

# 1. Introduction

Surface skinning or lofting is an approach used to construct surfaces from rows of data points constituting the boundaries of serial cross-sections of a given geometry [1-2]. The method is one of the most widely used tools for interactive shape description in computer-aided design [2]. With modern approaches to data capture, such as laser scanning, the use of the technique continues to grow, and it is now widely used in the construction of 3D computer-based surface models from scanned geometries. In CAD or CAGD, computationally efficient bi-parametric B-splines/NURBS representations have become the de-facto industry standard for smooth curves and surfaces [3-6]. As a result, the process of skinning or lofting using B-spline/NURBS curves and surfaces is an important problem [7-15].

Approaches to constructing a B-spline/NURBS surface to fit rows of data points can be broadly categorized into two types: approximation and interpolation. In approximation, the surface does not necessarily satisfy the given data precisely, but only approximately [7-11]. The approach is suitable when the number of rows of data points is large, or particularly when the data set contains noise. The technique involves fitting a family of approximating B-spline curves defined over a common knot vector. In interpolation, a surface is constructed that satisfies the given data precisely. This approach is suitable when the number of data points is not large and noise is not an issue [12- 15].

In either case, the technique can be implemented as follows:
(1) Each row can be fitted separately with an approximating or interpolating B-spline curve.
(2) The resulting curves can be made compatible via degree elevation and knot refinement to ensure that they all share a common knot vector [7].
(3) The B-spline surface can be generated by approximating or interpolating columns of resulting control points.

The above method is straightforward; however, it often results in a significant increase in the number of control points in the lofted B-spline surface. The problem is particularly profound where an interpolating surface is required. Solutions to date to mitigate this problem have focused on addressing the issue for surfaces comprised of open contours [7, 9,11-15]. Where an approximating surface will suffice, the number of control points required can be reduced [7-11]. Most of the work in this area has focused on the open contour case [7, 9, 11]. Where an interpolating surface is required, the problem is more complex. Most of the approaches to date in this area have also focused on the case of the open contour [12-15].

This paper focuses on closed B-spline curve and surface interpolation. We firstly examine the interpolation condition of the closed B-spline curve interpolation. That is the relationship between the parameter values and the domain knots. The key issue is what condition the parameter values and the domain knots should satisfy to guarantee the system matrix is invertible or full-rank. Then we apply the interpolation condition to lofted B-spline surface interpolation of serial closed contours.

The paper is laid out as follows. In Section 2, related work on open lofted B-spline surface interpolation and closed lofted B-spline surface approximation is described and discussed. Section 3 presents an overview of the necessary technical background. The problem summary is given in Section 4. In Section 5, the proposed conjectures and core contribution of the paper is presented. Then the related application of conjectures is described. Section 6 presents experimental results to demonstrate the applications of the related conjectures to the algorithms proposed to closed lofted B-spline surface interpolation. Finally, Section 7 concludes this paper.

# 2. Related work

Several researchers have developed feasible algorithms to B-spline curve interpolation and lofted B-spline surface interpolation. However, for lofted B-spline surface interpolation, much of the focus to date has been on skinning/lofting from open section curves. Where the case of closed section curves has been considered, only approximating techniques have been employed [8, 10].

## 2.1 Open B-spline curve interpolation

The widely used method for open B-spline curve interpolation is the so-called averaging technique [3-6], the knot can be computed via the average of the consecutive parameter values. With this method, the knots reflect the distribution of the corresponding parameter values.

## 2.2 Closed B-spline curve interpolation

For the closed B-spline curve interpolation, the averaging technique cannot be applied directly. The essential reason for this is the system matrix of closed B-spline curve interpolation introduces additional linear constraints. The earliest documented work we could find was that of Park and Kim [8]. It is a knot placement technique similar to the averaging technique referenced in section 2.1 for the open B-spline case. With that strategy, the corresponding knot vector reflects the distribution of parameter values. However, the adaptation presented for the closed case only works for odd degree B-splines. When the degree is even, it likely makes the system matrix ill-conditioned and leads to a bad quality of closed curve, and Park did discuss this point in his work [16]. At worst, a singular system matrix may occur. Park [16] addressed the choice of parameter values and knots in the closed B-spline curve interpolation by the *natural* [4] and *shifting* methods [16] for odd degree and even degree B-splines, respectively.

## 2.3 Lofting of open section curves using interpolation

When lofting serial B-spline curves to form a surface, various approaches have been considered to avoid the introduction of many additional control points. Piegl and Tiller [12] presented an open lofted B-spline interpolation approach that suggests giving the knot a certain degree of freedom so that each row of data points can be interpolated with as few new knots added as possible. Their approach chooses the knots for each B-spline curve interpolation from an input knot vector via the following steps. First, compute the anchor knot vector by averaging the parameter values, then they define an interval (a, b) to see if it consists of knots from the input knot vector. If yes, it selects the closest one; otherwise, the knot in the anchor knot vector will be used. To ensure numerical stability in their approach, the parameter values and knot vector satisfy the Schoenberg-Whitney condition [4].

Park [13] proposed another algorithm to reduce the number of control points in the case of open lofted surface interpolation. The approach first determines a common knot vector composed of fewer knots which contain enough degrees of freedom to guarantee the existence of a B-spline curve interpolating each row of data points. Although the approach generates an under-determined system, the unknown control points could be found by solving a strain energy minimization problem with linear constraints. Since the approximated strain energy form is a quadratic functional, the minimization problem leads to that of solving a system of equations by means of the Lagrange multiplier. This method produces a sequence of compatible B-spline curves that passing through corresponding rows of data points. Experimental results showed that Park's approach is more effective at reducing the control points than Piegl and Tiller's.

Wang et al. [14] pointed out the common knot vector determined via Park's method [13] cannot guarantee the existence of interpolating B-spline curves, as the resulting interpolating matrix may be rank-deficient. Unwanted wiggle or undulations may occur in the resulting lofted B-spline surface. To overcome the limitations of Park's approach, Wang et al. proposed a necessary and sufficient condition called the generalization of the Schoenberg-Whitney condition that can guarantee the existence of B-spline curves for each row interpolation. Using their algorithm, the common knot vector can guarantee the interpolating matrix is full-rank. Practical experiments showed that their algorithm could reduce more control points than Piegl and Tiller's approach while avoiding undesirable undulations.

Park [15] also presented an approach to open lofted surface interpolation based on universal parametrization [18-19]. This approach first finds the highest index of rows of data points, and then defines a common knot vector using the following steps: Compute the parameter values via chordal length parametrization [3] for rows with the highest index; Average these parameter values and determine the domain knots for degree $p$ via averaging technique [3-6]; Then, each B-spline curve can be created on a knot vector that is a subset of predefined common knot vector by adopting universal parametrization technique; Lastly, all the interpolating B-spline curves can be made compatible by inserting complementary knots into each knot vector. As the number of control points of each compatible B-spline curve equals the highest number of rows of data points, it leads to a compact representation of the lofted B-spline surface.

## 2.4 Lofting of closed contours based on approximation

Closed section curves described by periodic B-spline curves are needed in many practical applications for the creation of tubular surfaces (see Fig. 5 - Fig. 11). The early work of Park and Kim [8] addressed the problem of closed lofted B-spline approximation. In their approach, a bi-cubic B-spline surface closed in the contour direction and open in the longitudinal direction is constructed to approximate the predefined cross-sections by least-squares [3-4]. This technique can be called multiple curves approximation based on a common knot vector. In order to approximate a family of intermediate section curves with a common knot vector, the following procedure is employed:

(1) For the *i*-th contour, compute the domain knots according to the relationship (equal, less or greater) of the number of control points and the number of data points;
(2) Average all the domain knots and then use it to define a common knot vector.

Park and Kim [8] also introduced a binary search algorithm to reduce the number of knots in the common knot vector. By the iterative process, their method provides a smooth and accurate model for serial cross-sections.

Besides, when the number of data points less than the number of control points, the problem is underspecified. Thus the unwanted wiggle or undulation may occur in the resultant curve. A simple method for avoiding this defect in Ref [8] is adding some extra data points and make the problem well-defined or over-specified. Another approach to overcome the above shortcoming was proposed in the work of Park et al. [10]. Namely, B-spline curve approximation based on the least-squares and energy minimization.

Due to the existing methods [8, 10] for lofting of a series of closed contours is an approximation, rather than interpolation, so this paper firstly proposed three conjectures, then the existing algorithm of lofted B-spline surface interpolation to sequential open section curves can be extended to closed B-spline surface interpolation. All the details are organized in Section 5.

# 3. Technical background

## 3.1 Open B-spline curve and surface

We assume that the reader is familiar with the concepts of B-spline curves [3-6]. For notational convenience, we introduce the definition of B-splines. An open B-spline curve of degree $p$ is defined by a linear combination of B-spline functions.

$$\mathbf{C}(u) = \sum_{i=0}^{n} N_{i,p}(u)\mathbf{P}_i \qquad u \in [u_p, u_{n+1}] \tag{1}$$

where $\mathbf{P}_i$ are the control points and $N_{i,p}(u)$ are the normalized basis functions defined over a clamped knot vector

$$\mathbf{U}_{\text{clamped}}^{n,p}(u_i) = \{\underbrace{u_0, \cdots, u_p}_{p+1}, u_{p+1}, \cdots, u_n, \underbrace{u_{n+1}, \cdots, u_{n+p+1}}_{p+1}\} \; (u_i \leq u_{i+1}), \text{ where } u_0 = \cdots = u_p = 0 \quad u_{n+1} = \cdots = u_{n+p+1} = 1$$

As an extension of a parametric B-spline curve, a bi-parametric B-spline surface can be defined in tensor product form. Namely,

$$\mathbf{S}(u,v) = \sum_{i=0}^{m}\sum_{j=0}^{n} N_{i,p}(u)N_{j,q}(v)\mathbf{P}_{i,j} \; u \in [u_p, u_{m+1}] \quad v \in [v_q, v_{n+1}] \tag{2}$$

where $\mathbf{P}_{i,j}$ are the control net in 3D and $N_{i,p}(u), N_{j,q}(v)$ are the B-spline basis functions defined on the clamped knot vector $\mathbf{U}_{\text{clamped}}^{m,p}(u_i)$ and $\mathbf{V}_{\text{clamped}}^{n,q}(v_i)$, respectively.

## 3.2 Closed B-spline curve and surface

Similar to the open B-spline curve, a closed *p*-th degree B-spline curve [5, 8, 10] is defined as follows:

$$\tilde{\mathbf{C}}(u) = \sum_{i=0}^{n+p} N_{i,p}(u)\mathbf{P}_{i \bmod (n+1)} \qquad u \in [u_0, u_{n+1}] \tag{3}$$

where $N_{i,p}(u)$ are the normalized basis functions that defined on a called cyclic knot vector

$$\mathbf{U}_{\text{cyclic}}^{n,p}(u_i) = \{\underbrace{u_{-p}, \cdots, u_{-1}}_{p}, u_0, \cdots, u_{n+1}, \underbrace{u_{n+2}, \cdots, u_{n+p+1}}_{p}\}, \text{ where}$$

$$\begin{cases} u_{-i} = u_{-(i-1)} + u_{n-i+1} - u_{n-i+2} \\ u_0 = 0 \quad u_{n+1} = 1 \\ u_{n+i+1} = u_{n+i} + u_i - u_{i-1} \quad (i = 1, \cdots, p) \end{cases} \tag{4}$$

The values $u_i$ are the elements of a knot vector that is a non-decreasing sequence of real numbers and the knots $u_i \; (i = 0, \cdots, n+1)$ are called *domain knots*. The previous definitions ensure that the B-spline curve is closed, i.e. $\tilde{\mathbf{C}}(u_0) = \tilde{\mathbf{C}}(u_{n+1})$ and defined for $u \in [u_0, u_{n+1}]$ with a $C^{p-1}$ continuity everywhere.

Analogous to open B-spline surface, the (*p* x *q*)-th degree B-spline surfaces which are closed in the contour direction and open in the longitudinal direction can be expressed as follows:

$$\tilde{\mathbf{S}}(u,v) = \sum_{i=0}^{m}\sum_{j=0}^{n+q} N_{i,p}(u)N_{j,q}(v)\mathbf{P}_{i,\, j \bmod (n+1)} \quad u \in [u_p, u_{m+1}] \quad v \in [v_0, v_{n+1}] \tag{5}$$

where $\mathbf{P}_{i,j}$ are the control net in 3D and $N_{i,p}(u), N_{j,q}(v)$ are the B-spline basis functions defined on the clamped knot vector $\mathbf{U}_{\text{clamped}}^{m,p}(u_i)$ and the cyclic knot vector $\mathbf{V}_{\text{cyclic}}^{n,q}(v_i)$, respectively.

## 3.3 Open B-spline curve and surface interpolation

### 3.3.1 Open B-spline curve interpolation

Given a set of data points $\mathbf{Q}_0, \cdots, \mathbf{Q}_n$, and assuming that no two neighboring data points are the same. An open B-spline curve $\mathbf{C}(u)$ is sought to pass through these data points at certain parameter values $\mathbf{T}_{\text{open}} = \{t_0, \cdots, t_n\}$ and a clamped knot vector $\mathbf{U}_{\text{clamped}}^{\hat{n},p}$, which can be expressed in matrix form:

$$\mathbf{N}_{\text{open}} \mathbf{P} = \mathbf{Q} \tag{6}$$

where $\mathbf{P}_{(\hat{n}+1) \times 1} = \begin{bmatrix} \mathbf{P}_0 & \cdots & \mathbf{P}_{\hat{n}} \end{bmatrix}^T$ are the control points and $\mathbf{N}_{\text{open}}$ called system matrix, which is a $(n+1) \times (\hat{n}+1)$ matrix of scalars in the following form:

$$\mathbf{N}_{\text{open}} = \begin{bmatrix} N_{0,p}(t_0) & \cdots & N_{\hat{n},p}(t_0) \\ N_{0,p}(t_1) & \cdots & N_{\hat{n},p}(t_1) \\ \vdots & \ddots & \vdots \\ N_{0,p}(t_n) & \cdots & N_{\hat{n},p}(t_n) \end{bmatrix} \tag{7}$$

The parameter values $\mathbf{T}_{\text{open}}$ can be determined via the general exponent form [3-6].

### 3.3.2 Schoenberg-Whitney condition and its generalization

If $n = \hat{n}$, the system matrix is square and the above problem Eq. (6) leads to the well-known curve interpolation. The Schoenberg-Whitney condition [4] can guarantee the corresponding system matrix $\mathbf{N}_{\text{open}}$ invertible and gives each knot some flexibility. Namely, $\mathbf{U}_{\text{clamped}}^{n,p}$ and $\mathbf{T}_{\text{open}}$ satisfy the following condition:

$$t_{i-p-1} < u_i < t_i \quad (i = p+1, \cdots, n)$$

The averaging technique [3-6] works well for this case, which determines the knot vector via the following formula:

$$\begin{cases} u_0 = \cdots = u_p = 0 \quad u_{n+1} = \cdots = u_{n+p+1} = 1 \\ u_i = \frac{1}{p} \sum_{k=i-p}^{i-1} t_k \quad (i = p+1, \cdots, n) \end{cases} \tag{8}$$

The knots reflect the distribution of parameter values $t_i$ via this method. Furthermore, the system matrix $\mathbf{N}_{\text{open}}$ is totally positive and banded with a semi-bandwidth less than *p* (see Ref [4]). Hence, the linear system can be solved by Gaussian elimination without pivoting.

If $n < \hat{n}$, the system matrix is not square and the problem is underspecified. There may exist multiple open B-spline curves that pass through the given data points $\mathbf{Q}_i \ (i = 0, \cdots, n)$. Wang et al [14] pointed out that to guarantee the existence of the solution of Eq. (6), the system matrix $\mathbf{N}_{\text{open}}$ must be *full-rank*. To guarantee this point, the generalization of the Schoenberg-Whitney condition was proposed in their work. That is,

The system matrix $\mathbf{N}_{\text{open}}$ is a *full-rank* matrix if and only if there exists a clamped knot vector $\bar{\mathbf{U}} = \{\bar{u}_0, \bar{u}_1, \cdots, \bar{u}_{n+p+1}\}$, $\bar{u}_i \in \mathbf{U}_{\text{clamped}}^{\hat{n},p}$ such that $\mathbf{T}_{\text{open}}$ and $\bar{\mathbf{U}}$ satisfy the Schoenberg-Whitney condition.

### 3.3.3 B-spline interpolation based on energy minimization

Although Eq. (6) generates an under-determined system when $n < \hat{n}$, Park [13] suggested to compute the corresponding interpolation curve via minimizing the energy of the curve, which is defined as follows [10, 13]:

$$E(\mathbf{C}(u)) = \int_u \left( \alpha \|\dot{\mathbf{C}}(u)\|^2 + \beta \|\ddot{\mathbf{C}}(u)\|^2 \right) du = \mathbf{P}^T \mathbf{K} \mathbf{P} \tag{9}$$

where

$$\begin{cases} \mathbf{N}^{(k)} = \left[ N_{0,p}^{(k)}(u) \quad \cdots \quad N_{\hat{n},p}^{(k)}(u) \right]^T \\ \mathbf{P} = \left[ \mathbf{P}_0 \quad \cdots \quad \mathbf{P}_{\hat{n}} \right]^T \\ \mathbf{K} = \int_u \alpha \left( \mathbf{N}^{(1)} \mathbf{N}^{(1)T} \right) du + \int_u \beta \left( \mathbf{N}^{(2)} \mathbf{N}^{(2)T} \right) du \end{cases} \tag{10}$$

Here, $\mathbf{N}^{(k)}$ is a vector that storing the $k$-th derivatives of B-spline basis functions. $\mathbf{K}$ is called stiffness matrix, which is a positive definite, symmetric ($\mathbf{K} = \mathbf{K}^T$) and banded matrix with a bandwidth ($2p + 1$) and can be computed via the Gauss quadrature or analytic approach that proposed by Moore and Molloy [20]. The value $\alpha$ and $\beta$ are called stretching and bending coefficients, respectively. In this paper, we set $\alpha = 1.0$ $\beta = 0.2$, which are the same as Park's paper [10, 13].

Using the Lagrange multipliers $\mathbf{v}$ and linear constraints Eq. (6) to minimize $E(\mathbf{C}(u))$ leads to the following linear system.

$$\begin{bmatrix} \mathbf{K}_{(\hat{n}+1) \times (\hat{n}+1)} & \mathbf{N}_{\text{open}\ (\hat{n}+1) \times (n+1)}^T \\ \mathbf{N}_{\text{open}(n+1) \times (\hat{n}+1)} & \emptyset_{(n+1) \times (n+1)} \end{bmatrix} \begin{bmatrix} \mathbf{P} \\ \mathbf{v} \end{bmatrix} = \begin{bmatrix} \emptyset \\ \mathbf{Q} \end{bmatrix} \tag{11}$$

Consequently, the control points $\mathbf{P}_{(\hat{n}+1) \times 1}$ can be achieved by solving the above linear system.

### 3.3.4 Lofted B-spline surface interpolation to open section-curves

Given rows of data points $\mathbf{Q}_{\text{open}} = \{\mathbf{Q}_{i,j}\}$ $i = 0, \cdots, m; j = 0, \cdots, n_i$, where, $\mathbf{Q}_{i,j}$ denotes the $j$-th data point of the $i$-th row, $m + 1$ represents the number of rows and $n_i + 1$ corresponds to the number of points in the $i$-th row. The task of the lofted B-spline surface interpolation is to construct an open B-spline surface to exactly interpolate these points. The traditional method [2, 3] to this problem can be summarized in Algorithm -1:

**Algorithm 1**: steps for lofted B-spline surface interpolation to open section curves

(1) Interpolate the rows of data points with an open $p$-th degree B-spline curve in the following three steps:

  (1.1) compute the parameter values $\mathbf{T}^{(i)}$ of each row of data points.

  (1.2) compute the clamped knot vector $\mathbf{V}^{(i)}$ from $\mathbf{T}^{(i)}$ by the *averaging* technique.

  (1.3) the unknown control points can be obtained by solving a linear system, see Eq. (6)

(2) Merge the knot vector of each curve via knot refinement technique to make these open B-splines compatible, that is, all the curves are defined over a common knot vector $\mathbf{V}$.

(3) Interpolate a set of columns of control points to generate a lofted B-spline surface.

Although the traditional method is straightforward and easy-to-implement, it tends to result in an astonishing number of control points in the lofted B-spline surface. The situation will become more and more obvious as the increase of the number of section-curves. To overcome this issue, Piegl and Tiller proposed the algorithm that uses the existing knot, and the numerical stability problem, see Fig. 1, can be avoided via the Schoenberg-Whitney condition [4]. Wang et al. [14] used the generalization of the Schoenberg-Whitney condition to construct a common knot vector, which leads to all the system matrices are full-rank. As a consequence, the control points of the corresponding row can be achieved by minimizing the energy of the curve.

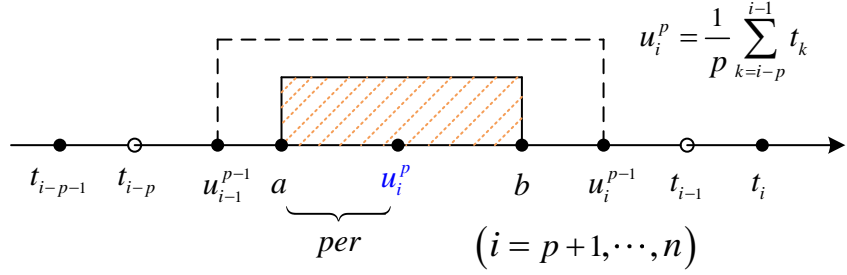

Fig. 1. The diagram of Piegl and Tiller's algorithm (using the existing knot)

## 3.4 Closed B-spline curve and surface interpolation

### 3.4.1 Closed B-spline curve interpolation

Given a set of data points $\mathbf{Q}_0, \cdots, \mathbf{Q}_n$, and assuming that no two neighboring data points are the same. A closed B-spline curve $\tilde{\mathbf{C}}(u)$ is sought to go through these data points at specific parameter values $\mathbf{T} = \{t_0, \cdots, t_n\}$ and a cyclic knot vector $\mathbf{U}_{\text{cyclic}}^{\hat{n},p}(u_i)$ that defined with Eq. (4), which leads to the following linear system:

$$\tilde{\mathbf{N}}\tilde{\mathbf{P}} = \tilde{\mathbf{Q}} \tag{12}$$

where $\tilde{\mathbf{P}}_{(\hat{n}+p+1)\times 1} = \begin{bmatrix} \mathbf{P}_0 & \cdots & \mathbf{P}_{\hat{n}} & \mathbf{P}_{\hat{n}+1} & \cdots & \mathbf{P}_{\hat{n}+p} \end{bmatrix}^{\mathrm{T}}$ are the control points and $\tilde{\mathbf{N}}$ is a $(n+p+1)\times(\hat{n}+p+1)$ system matrix of scalars in the following form:

$$\tilde{\mathbf{N}}_{(n+p+1)\times(\hat{n}+p+1)} = \begin{bmatrix} \mathbf{N} \\ \mathbf{A} \end{bmatrix} \quad \tilde{\mathbf{Q}}_{(n+p+1)\times 1} = \begin{bmatrix} \mathbf{Q} \\ \varnothing \end{bmatrix}$$

$$\mathbf{N}_{(n+1)\times(\hat{n}+p+1)} = \begin{bmatrix} N_{0,p}(t_0) & \cdots & N_{\hat{n}+p,p}(t_0) \\ N_{0,p}(t_1) & \cdots & N_{\hat{n}+p,p}(t_1) \\ \vdots & \ddots & \vdots \\ N_{0,p}(t_n) & \cdots & N_{\hat{n}+p,p}(t_n) \end{bmatrix} \tag{13}$$

$$\mathbf{A}_{p\times(\hat{n}+p+1)} = \begin{bmatrix} 1 & & & -1 & & \\ & \ddots & & & \cdots & & \ddots \\ & & 1 & & & & -1 \end{bmatrix}$$

$$\mathbf{Q}_{(n+1)\times 1} = [\mathbf{Q}_0 \quad \cdots \quad \mathbf{Q}_n]^{\mathrm{T}}$$

For notation convenience, we write the system matrix as $\tilde{\mathbf{N}}(n, \hat{n}, \mathbf{T})$. Similar to $\mathbf{T}_{\text{open}}$, the parameter values $\mathbf{T}$ of closed B-spline curves can be computed via the general exponent form [16]. That is,

$$t_0 = 0 \quad t_i = \frac{\sum_{j=0}^{i-1} |\mathbf{Q}_{(j+1)\bmod(n+1)} - \mathbf{Q}_j|^e}{\sum_{j=0}^{n} |\mathbf{Q}_{(j+1)\bmod(n+1)} - \mathbf{Q}_j|^e} \quad (i=1,\cdots,n) \tag{14}$$

If $n = \hat{n}$, the Eq. (12) degenerates to the interpolation problem. There exist three methods for closed B-spline curve interpolation. Namely, *averaging* [8], *natural* [4] and *shifting* method [16]. The *averaging* method [8] can be expressed as

$$u_i = (t_{i-1} + t_i + t_{i+1})/3 \quad (i = 1, \cdots, n) \quad \Leftarrow t_{n+1} = 1$$

The *natural* method [4], which sets the knots to coincide with the parameter values $\mathbf{T}$, that is, $u_i = t_i \quad (i = 1, \cdots, n)$

The *shifting* method determines the *domain knots* and parameter values $\tilde{\mathbf{T}} = \{\tilde{t}_0, \tilde{t}_1, \cdots, \tilde{t}_n\}$ in Park's work [16] as follows:

$$\begin{cases} u_0 = 0 \quad u_{n+1} = 1 \\ u_i = u_{i-1} + \begin{cases} (d_n + d_0)/2 & i = 1 \\ (d_{i-2} + d_{i-1})/2 & (i = 2, \cdots, n-1) \end{cases} \\ \tilde{t}_i = t_i + d_n/2 \quad (i = 0, \cdots, n) \end{cases} \quad (15)$$

where $d_i = t_{i+1} - t_i \quad (i = 0, \cdots, n) \quad \Leftarrow t_{n+1} = 1$

### 3.4.2 Lofted B-spline surface interpolation to serial closed contours

We firstly introduce the concept of contour alignment, which is a necessary step in the closed surface fitting. In general, the control net of a closed B-spline surface that closed in the contour direction consists of a set of closed control points. Each distinct alignment of the closed control points gives a different control net and subsequently results in different surface shape, as shown in Fig. 2. Therefore, for closed lofted B-spline surface interpolation, it is necessary to align the rows of data points before interpolating them. That means that the starting points should be appropriately positioned to avoid the resulting lofted surface twisting in the lofting direction. The starting points of each contour generate a baseline [8], which denotes the consecutive points set that traverse all the rows from the first to last in order. Intuitively, the resulting surface will twist in shape if the baseline is twisted heavily along the longitudinal direction.

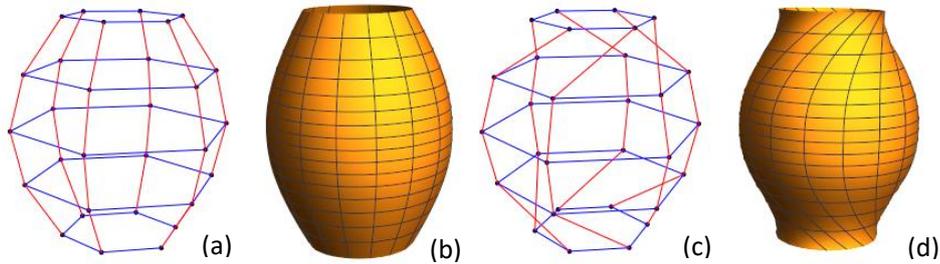

Fig. 2. An impact of distinct alignment of the control net

(a) original control net; (b) original B-spline surface; (c) control net via different alignment; (d) B-spline surface defined on the control net in (c)

Given a series of closed contours $\mathbf{Q} = \{\mathbf{Q}_{i,j}\} \quad i = 0, \cdots, m; \ j = 0, \cdots, n_i$. The task of the lofted B-spline surface interpolation is to construct a B-spline surface (closed in the $v$ direction and open in the $u$ direction) to exactly interpolate these points. The traditional approach, although we haven't find the related steps in existing literature, is organized in Algorithm -2:

---
**Algorithm 2**: steps for lofted B-spline surface interpolation to serial closed contours

(1) Align the serial closed contours via some algorithm, i.e. finding a good baseline.
(2) Interpolate the rows of data points with a closed *p*-th degree B-spline curve in the following three steps:

(2.1) compute the parameter values $\mathbf{T}^{(i)}$ of each row of data points.

(2.2) compute the cyclic knot vector $\mathbf{V}^{(i)}$ from $\mathbf{T}^{(i)}$ by the *natural* or *shifting* [16] methods.

(2.3) the unknown control points can be obtained by solving a linear system, see Eq. (12), built via $\mathbf{T}^{(i)}, \mathbf{V}^{(i)}$ or $\tilde{\mathbf{T}}^{(i)}, \mathbf{V}^{(i)}$.

(3) Clamp all the closed B-spline curves that defined over the cyclic knot vector.

(4) Merge the knot vector of each curve via knot refinement technique to make these closed B-splines compatible. That is, all the curves are defined over a common knot vector $\mathbf{V}$.

(5) Interpolate a set of columns of control points to generate a lofted B-spline surface.

Note that step-(3) is important and necessary due to all the closed B-spline curves are defined with a cyclic knot vector. To make them compatible, we should clamp them to clamped knot vector one by one before knot refining. In addition, the step-(2.3) pointed out that we must treat the degree in different ways, which is totally different from the lofted B-spline surface interpolation to serial open section curves.

Clamping is a technique that transforms a B-spline curve that defined over an unclamped knot vector to one defined on a clamped knot vector, which is used in making a series of closed B-spline curves defined over a cyclic knot vector compatible. Piegl and Tiller [3] pointed out there is no difference between clamped and unclamped B-spline curves from the perspective of the algorithm. A closed B-spline curve defined over a cyclic knot vector can be easily transformed into the one defined on a clamped knot vector, and vice versa. According to the original unclamping algorithm (see page 577 of Ref [3]) presented by Piegl and Tiller, we present its corollary, an algorithm that clamps a B-spline curve. The algorithm that clamping a B-spline curve that defined over a cyclic knot vector is outlined in Procedure 1.

Analogous to the traditional method of lofted B-spline surface interpolation to serial open section-curves, the closed case also leads to the explosion of the number of control points as the increase of rows.

**Listing 1**: Algorithm for transforming a closed B-spline curve to clamped form

Procedure 1 **clamp_curve**$(\mathbf{P}, \mathbf{U}, p)$

INPUT:

$\mathbf{P} = \{\mathbf{P}_0, \cdots, \mathbf{P}_n\}$, the control points

$\mathbf{U} = \{u_0, \cdots, u_{n+p+1}\}$, knot vector defined in unclamped style

$p$, the degree of B-spline curve

OUTPUT:

$\mathbf{P}, \mathbf{U}$ for clamped B-spline curve

ALGORITHM:

//on the left side of knot-vector

**for** $i = p - 2$ **to** $0$ **do** {
  **for** $j = 0$ **to** $i$ **do** {
    $\alpha = \dfrac{u_p - u_{p-1-i+j}}{u_{p+j+1} - u_{p-1-i+j}}$
    $\mathbf{P}_j = (1-\alpha)\mathbf{P}_j + \alpha \mathbf{P}_{j+1}$
  }
  $u_{p-i-1} = u_p$
}

//on the right side of knot-vector

**for** $i = p - 2$ **to** $0$ **do** {
  **for** $j = 0$ **to** $i$ **do** {
    $\alpha = \dfrac{u_{n+1} - u_{n-j}}{u_{n-j+i+2} - u_{n-j}}$
    $\mathbf{P}_{n-j} = (1-\alpha)\mathbf{P}_{n-j} + \alpha \mathbf{P}_{n-j-1}$
  }
  $u_{n+i+2} = u_{n+1}$
}

# 4. Problem summary

Although the open B-spline curve and closed B-spline curve are defined in the same mathematical formula, the latter needs to the last *p* control points coincide with the first *p* control points when the knot vector is cyclic, i.e.

$\mathbf{P}_{\hat{n}+i} = \mathbf{P}_i$ $(i=1,\cdots,p)$, which leads to the extra constraint conditions. Hence, the closed B-spline curve interpolation is totally different from the open B-spline curve interpolation. We also discover this point via the formula of system matrix $\mathbf{N}_{\text{open}}$ and $\tilde{\mathbf{N}}$, the latter has a constant block matrix $\mathbf{A}_{p\times(\hat{n}+p+1)}$. That means the method of knot determination and the related numerical stability condition of B-spline curve interpolation is not applicable to the closed case. Namely, the classic Schoenberg-Whitney condition and its generalization cannot be applied to the closed B-spline curve interpolation, and the related algorithms [12-14] cannot be extended to the lofted B-spline surface interpolation to a sequence of closed contours directly.

In addition, although the existing methods have addressed the choice of parameter values and domain knots in the closed B-spline curve interpolation by the *averaging*, *natural* and *shifting* methods, these three strategies determine the unique relationship between parameter values and knots and the interpolation condition like the Schoenberg Whitney condition [4] used in open case is not proposed. We denote this problem as **Problem-1**.

If $n < \hat{n}$, there may exist multiple closed B-spline curves that pass through the given data points $\mathbf{Q}_i$ $(i=0,\cdots,n)$. To guarantee that the existence of the solution of Eq. (12), the system matrix $\tilde{\mathbf{N}}_{(n+p+1)\times(\hat{n}+p+1)}$ must be *full-rank* [14]. However, according to our limited knowledge, there is no such theorem providing a condition to guarantee the system matrix $\tilde{\mathbf{N}}_{(n+p+1)\times(\hat{n}+p+1)}$ is full-rank in the existing literature. We denote this problem as **Problem-2**.

Due to the existence of **Problem-1** and **Problem-2**, Piegl and Tiller's algorithm [12] and Wang's method [14] are only feasible for open lofted B-spline surface interpolation. As a result, the lofting or skinning of serial closed contours using interpolating techniques remains a problem.

# 5. Proposed approach

In this paper, we proposed some useful conjectures to try to address the above issues, i.e. Problem-1 and Problem-2. We summarized it as follows:

## 5.1 Conjectures for system matrix of closed curve interpolation

**Conjecture 1:**

Given that the degree of B-spline is *p*, the *domain knots* of the cyclic knot vector $\mathbf{U}$ is $\bar{\mathbf{D}} = \{u_0,\cdots,u_{n+1}\}$ and the parameter values $\mathbf{T} = \{t_0, t_1, \cdots, t_n\}$, then the system matrix $\tilde{\mathbf{N}}(n, \hat{n}, \mathbf{T})$ for closed B-spline interpolation is **invertible** if the following condition holds:

1) If *p* is odd, the system matrix is $\tilde{\mathbf{N}}(n, \hat{n}, \mathbf{T})$, the corresponding condition for $\mathbf{T}$ and $\bar{\mathbf{D}}$ is:

$$\frac{t_{i-1}+t_i}{2} < u_i < \frac{t_i+t_{i+1}}{2} \quad (i=1,\cdots,n)$$

2) If *p* is even, the system matrix is $\tilde{\mathbf{N}}(n, \hat{n}, \tilde{\mathbf{T}})$, the relevant condition for $\tilde{\mathbf{T}}$ and $\bar{\mathbf{D}}$ is:

$$\tilde{t}_{i-1} < u_i < \tilde{t}_i \quad (i=1,\cdots,n)$$

where $\tilde{t}_i = t_i + \frac{d_n}{2}$ $(i=0,\cdots,n)$, see Eq. (15).

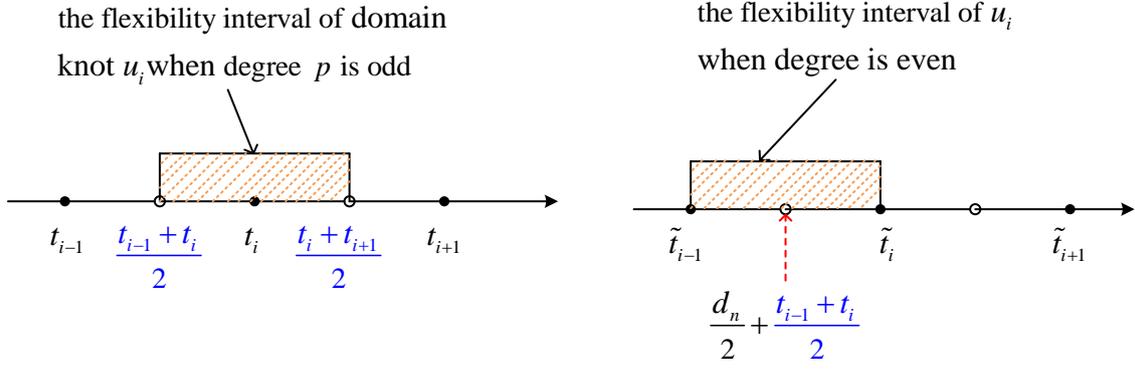

Fig. 3. The diagram of conjecture-1

**Conjecture 2:**

Suppose that the degree of B-spline curve $p$, the *domain knots* of the corresponding cyclic knot vector is $\mathbf{D}=\{u_0,\cdots,u_{\hat{n}+1}\}$ and the parameter values $\mathbf{T}=\{t_0,t_1,\cdots,t_n\}$, then the system matrix for closed B-spline curve interpolation is **full-rank** if there exist a *domain knots* $\bar{\mathbf{D}}=\{\bar{u}_0,\bar{u}_1,\cdots,\bar{u}_{n+1}\}$ $\bar{u}_i \in \mathbf{D}$ and $\mathbf{T},\bar{\mathbf{D}}$ (odd degree), or $\tilde{\mathbf{T}},\bar{\mathbf{D}}$ (even degree) satisfy the condition of **Conjecture 1**

Some comments are in order:
(1) The Conjecture 1 play the roles of the Schoenberg-Whitney condition of the closed B-spline curve interpolation.
(2) The Conjecture 2 is an analogy of the generalization of Schoenberg-Whitney condition, see Theorem-1 of Ref [14], which is the generalization of the Conjecture 1.
(3) For the Conjecture-1, if $u_i = t_i$, it degenerates to *natural* methods and if $u_i = \dfrac{t_{i-1}+t_i+t_{i+1}}{3}$, it degenerates to the *averaging* method. If $u_i = \left(\tilde{t}_{i-1}+\tilde{t}_i\right)/2$, it is the *shifting* method.

## 5.2 Applications to the existing lofted interpolation algorithms

In this paper, we use the chord-length parametrization to compute the parameter values $\mathbf{T}$. Because it is quite satisfactory and adequate in most practical applications [8, 9], and it is assumed that the closed contours have been aligned beforehand.

### 5.2.1 Adapts Piegl and Tiller's algorithm based on Conjecture 1

The Conjecture 1 gives the knot some flexibility similar to the Schoenberg-Whitney condition [4]. Based on the idea of Piegl and Tiller's algorithm [12], B-spline curve interpolation algorithm based on existing knot vector, shown in Fig. 1, we develop a closed B-spline curve interpolation method that attempts to choose the knots from a given input knot vector without causing the numerical stability problem. Summarized below are its overall steps.

To apply Piegl and Tiller's algorithm to closed B-spline curve interpolation based on the existent knot vector, we need to define two vectors, called *anchor domain knots* $\mathbf{D}^p$ and *bound-value vector* $\mathbf{X}$, which store the corresponding values as follows:

$$u_i^p = \begin{cases} t_i & p \text{ is odd} \\ \dfrac{d_n}{2}+\dfrac{t_{i-1}+t_i}{2} & p \text{ is even} \end{cases} (i=1,\cdots,n) \qquad (16)$$

$$u_0^p = 0 \quad u_{n+1}^p = 1$$

$$x_i = \begin{cases} \dfrac{t_i + t_{i+1}}{2} & p \text{ is odd} \\ \dfrac{d_n}{2} + t_i & p \text{ is even} \end{cases} (i = 0, \cdots, n) \qquad (17)$$

Because of $x_{i-1} < u_i^p < x_i$, analogous to Piegl and Tiller's [12] approach, we also define an interval (a, b), shown in Fig. 4, to give each knot some flexibility, where

$$\begin{cases} a = (1 - per) * u_i^p + per * x_{i-1} \\ b = (1 - per) * u_i^p + per * x_i \end{cases}$$

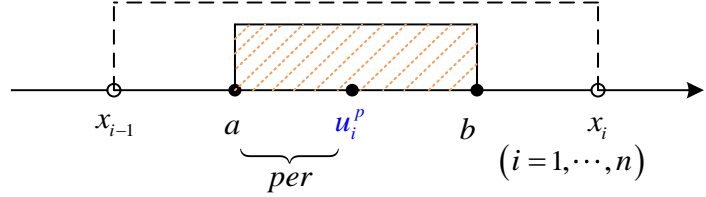

$(i = 1, \cdots, n)$

Fig. 4. The diagram of knot interval used in closed B-spline curve algorithm

The factor *per* denotes the percentage of interval $[x_{i-1}, x_i]$ use. The corresponding pseudocode as shown in Procedure 2.

**Listing 2**: Algorithm for closed B-spline curve interpolation based on a given knot vector

Procedure 2 **interpolate_points_by_input_knots**($\mathbf{Q}_i, \mathbf{T}, \bar{\mathbf{U}}, p, per$)

INPUT:

$\mathbf{Q}_i (i = 0, \cdots, n)$, data points    $\mathbf{T} = \{t_0, \cdots, t_n\}$, parameter values

$\bar{\mathbf{U}}$, input knot vector    $p$, required degree    *per*, percentage of interval use

OUTPUT:

$\tilde{\mathbf{R}}_i (i = 0, \cdots, n + p)$, the control points of closed B-spline curve

$\tilde{\mathbf{U}} = \{\underbrace{0, \cdots, 0}_{p}, \tilde{u}_0, \cdots, \tilde{u}_{n+1}, \underbrace{1.0, \cdots, 1.0}_{p}\}$, clamped knot vector for closed B-spline curve

$\bar{\mathbf{U}}$, updated input knot vector

ALGORITHM:

$\mathbf{D}^p = \{u_0^p, \cdots, u_{n+1}^p\} \leftarrow$ anchor domain knots determined by Eq.(16)

$\mathbf{X} = \{x_0, \cdots, x_n\} \leftarrow$ bound-value vector computed via Eq.(17)

$\mathbf{D} = \{u_0, \cdots, u_{n+1}\} \leftarrow$ memory for the domain knots

**for** $i = 1$ **to** $n$ **do** {

$\quad a = (1 - per) * u_i^p + per * x_{i-1}$

$\quad b = (1 - per) * u_i^p + per * x_i$

$\quad \bar{u}_k \leftarrow$ the knot in $\bar{\mathbf{U}}$ that closest to $u_i^p$

$\quad$ **if** $(\bar{u}_k \in (a,b))$ $u_i = \bar{u}_k$ **else** $u_i = u_i^p$

}

$\mathbf{U} \leftarrow$ cyclic knot vector based on $\mathbf{D}$ and Eq.(4)

$\mathbf{R}_i \leftarrow$ control points by interpolating the $\mathbf{Q}_i$ with $\mathbf{T}, \mathbf{U} / \tilde{\mathbf{T}}, \mathbf{U}$

$\tilde{\mathbf{R}}_i, \tilde{\mathbf{U}} \leftarrow$ clamp the closed curve via Procedure-1

$\bar{\mathbf{U}} \leftarrow \mathbf{merge}(\bar{\mathbf{U}}, \tilde{\mathbf{U}})$

Procedure 2 accepts the data points and a knot vector as input. It then judges whether the knot in a given knot vector can be used to interpolate the data points. Firstly, it sets up the anchor domain knots $\mathbf{D}^p$ for degree $p$ and another one called bound-value vector $\mathbf{X}$. If there are input knots in a flexibility interval, the algorithm chooses the one that nearest to the anchor domain knot $u_i^p$. Otherwise, the corresponding anchor knot $u_i^p$ is used. After the curve interpolation is done, the input knot vector $\bar{\mathbf{U}}$ is updated by adding the anchor knots whose flexibility interval did not consist of input knots.

**Listing 3**: Algorithm for interpolating a serial of rows of data points

Procedure 3 **interpolate_all_row_points**($\mathbf{Q}_{i,j}, q, per$)

INPUT:

$\mathbf{Q}_{i,j} (i = 0, \cdots, m; j = 0, \cdots, n_i)$, rows of data points

$q$, the degree in the row direction    $per$, percentage of interval use

OUTPUT:

$\mathbf{R}_{i,j} (i = 0, \cdots, m; j = 0, \cdots, n+q)$, control points of closed B-spline curves

$\bar{\mathbf{V}} = \{\underbrace{0, \cdots, 0}_{q}, \bar{v}_0, \cdots, \bar{v}_{n+1}, \underbrace{1.0, \cdots, 1.0}_{q}\}$, common knot vector

ALGORITHM:

$n_{\max} = \max_{i=0}^{m} n_i \leftarrow$ the highest index of rows of data points

compute the parameter of the row whose index equals to $n_{\max}$

$\mathbf{T} \leftarrow$ average these parameter values

$\mathbf{D} = \{\bar{v}_0, \cdots, \bar{v}_{n_{\max}+1}\} \leftarrow$ domain knots computed by $\mathbf{T}$ and natural/shifting method

$\bar{\mathbf{V}} = \{\underbrace{0, \cdots, 0}_{q}, \bar{v}_0, \cdots, \bar{v}_{n_{\max}+1}, \underbrace{1.0, \cdots, 1.0}_{q}\} \leftarrow$ initial input knot vector based on $\mathbf{D}$

**for** $i = 0$ **to** $m$ **do** {

$\mathbf{T}^{(i)} \leftarrow$ compute the parameter values of $\mathbf{Q}_{i,j} (j = 0, \cdots, n_i)$

$\left.\begin{array}{l}\tilde{\mathbf{R}}_{i,j} (j = 0, \cdots, n_i + q) \\ \mathbf{V}_i = \{\underbrace{0, \cdots, 0}_{q}, v_0, \cdots, v_{n_i+1}, \underbrace{1.0, \cdots, 1.0}_{q}\} \\ \bar{\mathbf{V}} \leftarrow \text{updated input knot vector}\end{array}\right\} \leftarrow \begin{array}{c}\textbf{interpolate\_points\_by\_input\_knots} \\ (\mathbf{Q}_i, \mathbf{T}^{(i)}, \bar{\mathbf{V}}, q, per)\end{array}$

}

$\bar{\mathbf{V}} = \{\underbrace{0, \cdots, 0}_{q}, \bar{v}_0, \cdots, \bar{v}_{n+1}, \underbrace{1.0, \cdots, 1.0}_{q}\} \leftarrow$ common knot vector of all B-spline curves

$\mathbf{R}_{i,j} (j = 0, \cdots, n+q) \leftarrow$ make all the curves compatible via knot refinement

The idea that passing a knot vector into the closed B-spline curve interpolation procedure is very important when interpolating a sequence of rows of data points. That is, each row of data points is interpolated independently with a given knot vector passed in for each row and then updated subsequently. It leads to that each B-spline curve owns many knots in common and the explosion of the number of the common knot vector can be avoided when making these curves compatible.

The overall procedure is summarized in Procedure-3. The value of the variable *per* is restricted in the interval [0, 1], represents the flexibility of a knot. A larger *per* could decrease the number of control net. In addition, when the factor *per* equals to 0, this algorithm degenerates to the traditional lofting method [2, 3].

### 5.2.2 Adapts Park and Wang's algorithm based on Conjecture 2

The key step of Park's approach is constructing a common knot vector. For the common cyclic knot vector, we just need to determine a common domain knots. Although this method generates an under-determined system, see Eq. (12), the corresponding control points $\tilde{\mathbf{P}}_{(\hat{n}+p+1)\times 1}$ of closed B-spline curve interpolation can be achieved via minimizing the energy of the curve. That leads to solving the below linear system:

$$\begin{bmatrix} \tilde{\mathbf{K}}_{(\hat{n}+p+1)\times(\hat{n}+p+1)} & \tilde{\mathbf{N}}^{\mathbf{T}}_{(\hat{n}+p+1)\times(n+p+1)} \\ \tilde{\mathbf{N}}_{(n+p+1)\times(\hat{n}+p+1)} & \varnothing_{(n+p+1)\times(n+p+1)} \end{bmatrix} \begin{bmatrix} \tilde{\mathbf{P}} \\ \tilde{\mathbf{v}} \end{bmatrix} = \begin{bmatrix} \varnothing \\ \tilde{\mathbf{Q}} \end{bmatrix} \quad (18)$$

where $\tilde{\mathbf{K}}_{(\hat{n}+p+1)\times(\hat{n}+p+1)}$ is the stiffness matrix defined in Eq. (10) with a cyclic knot vector. $\tilde{\mathbf{v}}_{(n+p+1)\times 1}$ is a vector storing Lagrange multiplier and $\tilde{\mathbf{N}}$ is determined via parameter values $\mathbf{T}$ and $\tilde{\mathbf{T}}$ for odd degree and even degree, respectively.

**Listing 4**: Algorithm for building the domain knots from given domain knots

Procedure 4 **build_domain_knots_by_input_knots**($\mathbf{T}, \mathbf{D}, p, per$)

INPUT:

$\mathbf{T} = \{t_0, \cdots, t_n\}$, parameter values of $\mathbf{Q}_i$     $\mathbf{D} = \{u_0, \cdots, u_{\hat{n}}, u_{\hat{n}+1}\}$, input domain knots

$p$, required degree     $per$, percentage of interval use

OUTPUT:

$\bar{\mathbf{D}} = \{\bar{u}_0, \cdots, \bar{u}_n, \bar{u}_{n+1}\}$, constructed domain knots for contour points $\mathbf{Q}_i$

ALGORITHM:

$\mathbf{D}^p = \{u_0^p, \cdots, u_{n+1}^p\} \leftarrow$ anchor domain knots determined by Eq.(16)

$\mathbf{X} = \{x_0, \cdots, x_n\} \leftarrow$ bound-value vector computed via Eq.(17)

$\bar{\mathbf{D}} = \{\bar{u}_0, \bar{u}_1, \cdots, \bar{u}_n, \bar{u}_{n+1}\} \leftarrow$ memory for the domain knots

$span = 0$

**for** $i = 1$ **to** $n$ **do** {

  $a = (1 - per) * u_i^p + per * x_{i-1}$

  $b = (1 - per) * u_i^p + per * x_i$

  **while** ($u_{i+span} < a$) {

    $span++$

    **if** $(i + span > \hat{n} + 1)$ **break** the **for**-loop

  }

  **if** $(u_{i+span} < b)$ $\bar{u}_i = u_{i+span}$ **else** { $\bar{u}_i = u_i^p$ **and** $span$-- }

}

**for** $k = i$ **to** $n$ **do** { $\bar{u}_k = u_k^p$ } $\leftarrow$ the remaining domain knots

**return** $\bar{\mathbf{D}}$

Similar to Wang's [14] (see Theorem-1 and Procedure-2 of that paper) method that constructing a knot vector that guarantees the corresponding system matrix is *full-rank*, we use **Conjecture-2** to build a domain knots that make the system matrix of closed B-spline curve interpolation is full-rank. The related steps are summarized in Procedure 4. Furthermore, unlike the method of Piegl and Tiller [12], we notice that the clamping and knot refinement is not needed in Park and Wang's approach.

The common domain knots can be determined by the following two steps: first, an initial common domain vector $\mathbf{D}$ is computed, then for each row, a domain vector $\mathbf{D}^{(i)}$ is obtained from Procedure 4 and $\mathbf{D}$ is updated by merging itself with $\mathbf{D}^{(i)}$. The algorithm is described as follows.

**Listing 5**: Algorithm for building common domain knots of rows of data points

Procedure 5  **build_common_domain_knots**($\mathbf{Q}_{i,j}, q, per$)
INPUT:
$\mathbf{Q}_{i,j}(i = 0, \cdots, m; j = 0, \cdots, n_i)$, rows of data points
$q$, required degree of row    $per$, percentage of interval use
OUTPUT:
$\mathbf{D} = \{u_0, \cdots, u_{\hat{n}}, u_{\hat{n}+1}\}$, constructed common domain knots
ALGORITHM:
$n_{max} = \max_{i=0}^{m} n_i \leftarrow$ the highest index of rows of data points
compute the parameters of the row whose index equals to $n_{max}$
$\mathbf{T} \leftarrow$ average these parameter values
$\mathbf{D} = \{u_0, \cdots, u_{n_{max}}, u_{n_{max}+1}\} \leftarrow$ initial domain knots computed by $\mathbf{T}$ and Eq.(16)
**for** $i = 0$ **to** $m$ **do** {
    $\mathbf{T}^{(i)} \leftarrow$ compute the parameter values of $\mathbf{Q}_{i,j}(j = 0, \cdots, n_i)$
    $\mathbf{D}^{(i)} \leftarrow$ **build_domain_knots_by_input_knots**($\mathbf{T}^{(i)}, \mathbf{D}, q, per$)
    $\mathbf{D} \leftarrow$ **merge**($\mathbf{D}, \mathbf{D}^{(i)}$)
}
**return D**

Once the common domain knots is built, we can use the linear system Eq. (12) and the common *domain knots* to compute all the control points of the B-spline curves that pass through all the rows of data points $\mathbf{Q}_{i,j}$. Lastly, the lofted B-spline surface can be achieved via interpolating a set of columns of control points.

# 6. Experimental results

The proposed approach (Procedure 1-5) has been implemented in C and Wolfram Language, and tested for various sets of closed rows of data points. In this paper, a (*p* x *q*)-th B-spline surfaces (closed in the row direction and open in the column direction, *p* = 3, *q* = 2 to 6) were used and chordal-length parametrization was applied to interpolate a sequence of control points in the column direction. In the following, seven groups of point-sets are included to demonstrate the usability and quality of the related algorithm. All the baselines used for contour alignment are denoted in red. In addition, Fig. 5 - Fig. 8 were generated by Piegl and Tiller's approach and Fig. 9 – Fig. 12 were generated via Park and Wang's method.

Firstly, Piegl and Tiller's and Park and Wang's approach are applied to five models that the number of rows is not large, respectively. The first one is a human jaw, shown in Fig. 5, nine rows of data points were scanned from an entity model. The number of points in each row ranges from 111 to 136. The latter two are blades of aero engine, shown in Fig. 6 – Fig. 7, the corresponding number of points of each row ranges from 140 to 160 and 106 to 110, respectively. The last two model are wineglass and water bottle, shown in Fig. 8 – Fig. 9, we sampled eight and twelve rows of data points from their surface.

The relevant number of points of each row ranges from 17 to 54 and 54 to 80, respectively. In this paper, we apply different degree in the row direction to five models and the factor *per* equals to 1.0. The related results and information are summarized in Table 3.

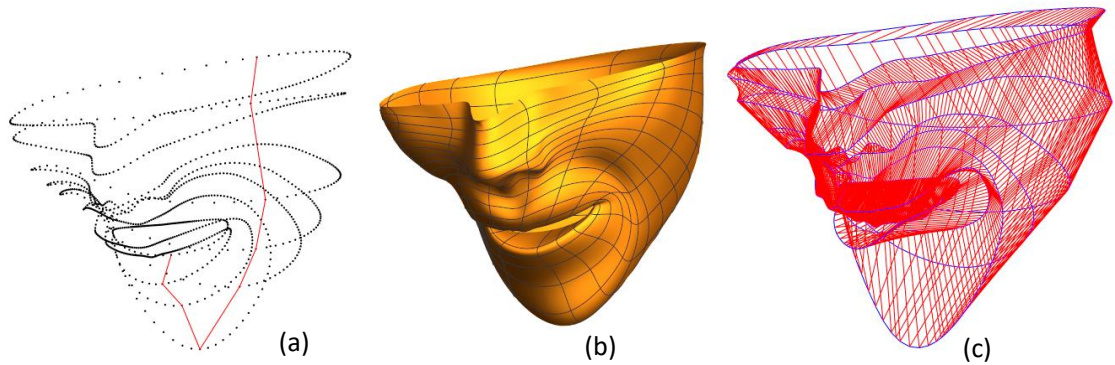

(a)　　　　　　　　　　　　(b)　　　　　　　　　　　　(c)

Fig. 5. Lofted B-spline surface to rows of data points sampled from a jaw model

(a) sampled points and based line; (b) lofted B-spline surface; (c) control net

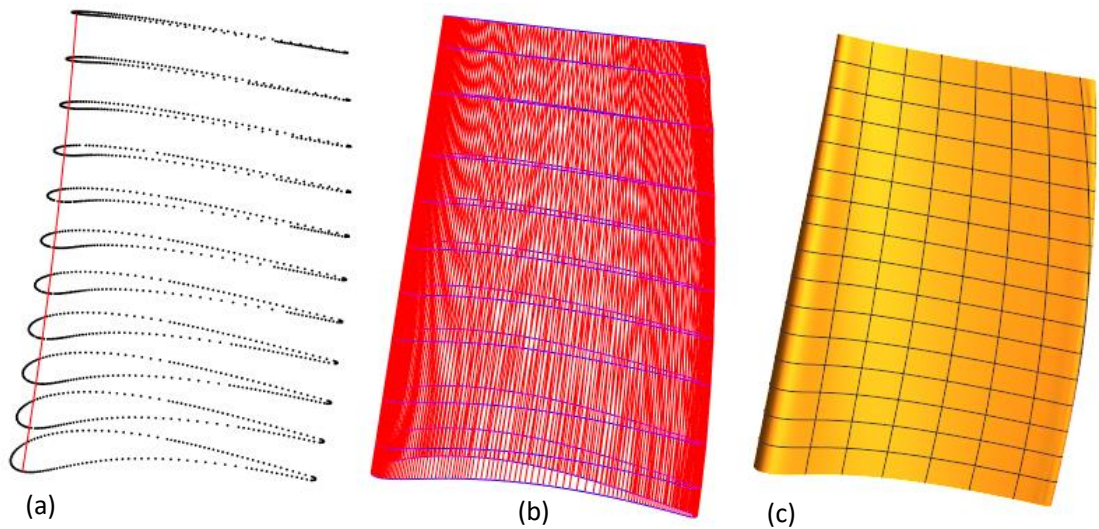

(a)　　　　　　　　　　　　(b)　　　　　　　　　　　　(c)

Fig. 6. Lofted B-spline surface to rows of data points sampled from a blade

(a) sampled points and based line; (b) control net; (c) lofted B-spline surface

**Table 3**

The dimension of control net of different models with *per* = 1.0

| model | min $n_i$ | max $n_i$ | v-degree ($q$) | control net(Piegl) | control net(Park) |
|---|---|---|---|---|---|
| jaw | 110 | 135 | 2 | 9 x 190 | 9 x 190 |
| blade-A | 139 | 159 | 3 | 11 x 222 | 11 x 221 |
| blade-B | 105 | 109 | 4 | 5 x 117 | 5 x 117 |
| wineglass | 18 | 53 | 5 | 8 x 89 | 8 x 79 |
| bottle | 53 | 79 | 6 | 12 x 89 | 12 x 92 |

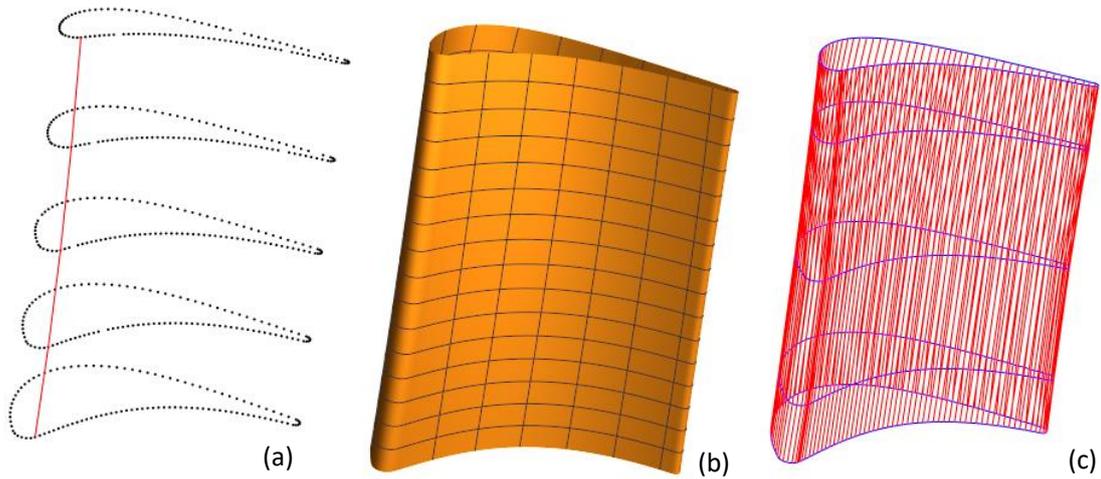

Fig. 7. Lofted B-spline surface to rows of data points sampled from a blade
(a) sampled points and based line; (b) lofted B-spline surface; (c) control net;

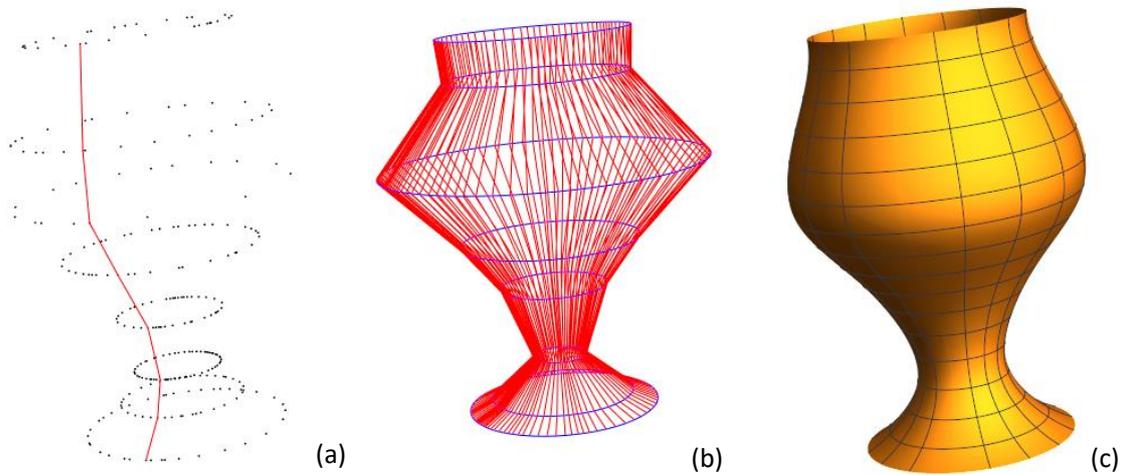

Fig. 8. Lofted B-spline surface to rows of data points sampled from a wineglass
(a) sampled points and based line; (b) control net; (c) lofted B-spline surface;

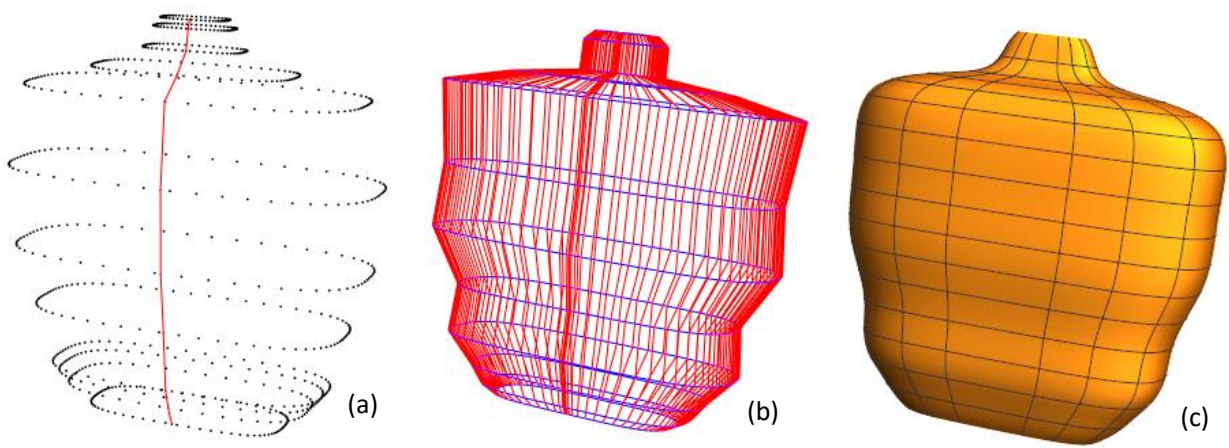

Fig. 9. Lofted B-spline surface to rows of data points sampled from a water bottle
(a) sampled points and based line; (b) control net; (c) lofted B-spline surface.

Secondly, 92 rows of data points were scanned from the surface of a human head. For the convenience of the display, only 46 rows are shown in Fig. 10-(a). Piegl and Tiller's algorithm and Park and Wang's approach were tested for six different row numbers and four different scale factors (0.25 to 1). The quantitative comparison is summarized in Table 4. Fig. 10 shows the application of the proposed method to 92 rows of data points with *per* = 1.0, where the number of points for each row ranges from 91 to 180. Piegl and Tiller's and Park and Wang's approaches require 92 x 182 control points, shown in Fig.10-(f), to interpolate these rows.

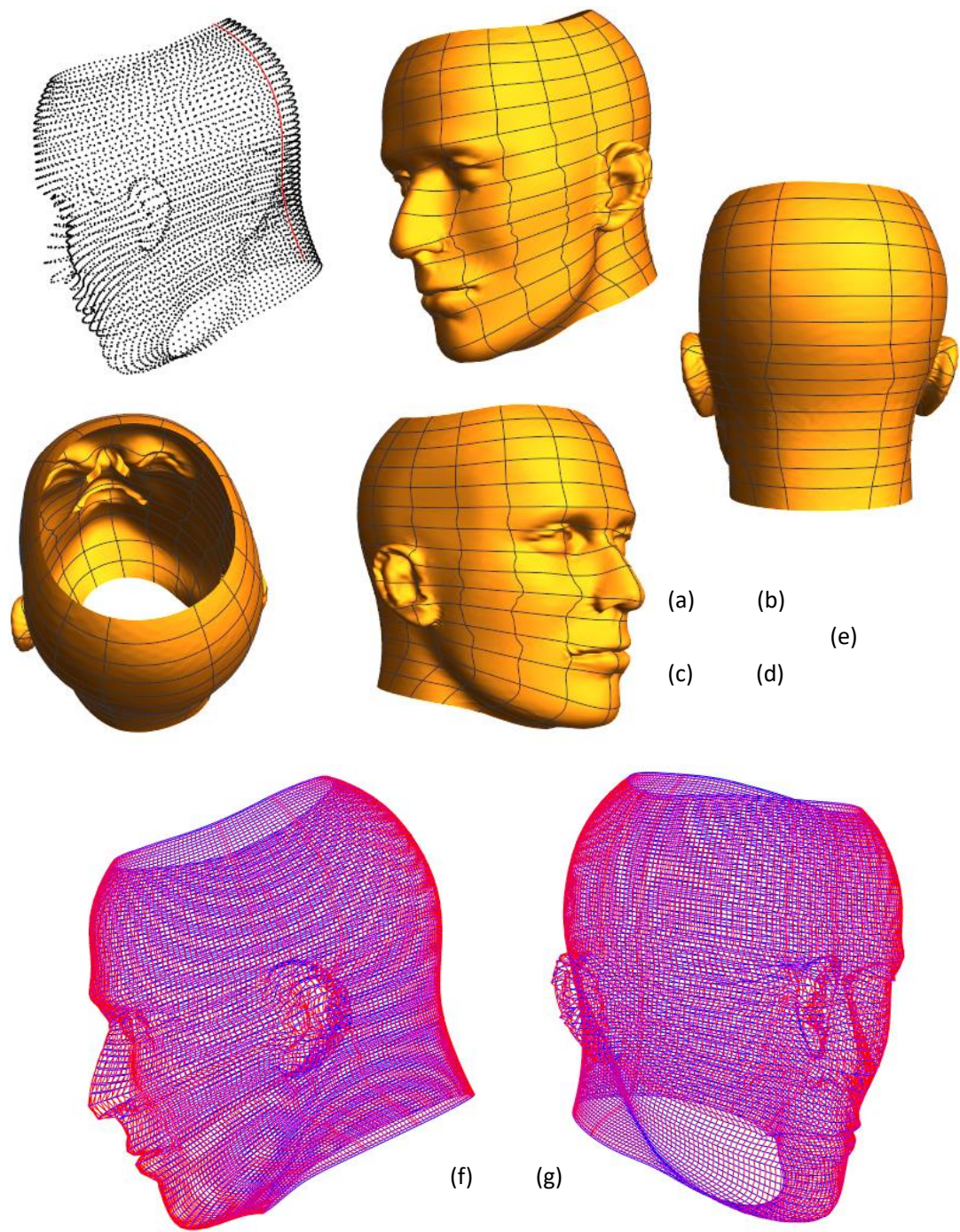

Fig. 10. Lofted B-spline surface to rows of data points sampled from a human head
(a) sampled points and based line; (b) lofted B-spline surface with *per*=1.0; (c) - (e) surfaces at different angles; (f) - (g) control net (92 x 182)

Table 4
Comparison of the number of control points for human head by different number of rows

| *m* | | 5 | 10 | 20 | 40 | 80 | 91 |
|---|---|---|---|---|---|---|---|
| min $n_i$ | | 91 | 91 | 91 | 91 | 90 | 90 |
| max $n_i$ | | 98 | 98 | 100 | 179 | 179 | 179 |
| | | | | | | | |
| Piegl's approach (*q*=3) | *per*=1.00 | 101 | 101 | 103 | 182 | 182 | 182 |
| | *per*=0.75 | 131 | 161 | 181 | 349 | 354 | 354 |
|  | per=0.50 | 169 | 211 | 230 | 424 | 435 | 435 |
|  | per=0.25 | 220 | 307 | 384 | 767 | 836 | 836 |
| Park's approach (q=3) | per=1.00 | 101 | 101 | 103 | 182 | 182 | 182 |
|  | per=0.75 | 131 | 161 | 181 | 346 | 353 | 353 |
|  | per=0.50 | 169 | 211 | 230 | 423 | 434 | 434 |
|  | per=0.25 | 220 | 307 | 383 | 772 | 835 | 835 |

Thirdly, we apply the above two methods to a liver model. A set of cross-sections are generated from the iso-curves of a liver model. Then the rows of data points can be obtained via these cross-sections. We test all approaches for five different row numbers and four different scale factors. Table 5 summarizes the related results. In this case, 48 rows of data points were used to demonstrate the resulting surface of the liver model, shown in Fig. 11, with per = 1.0. In addition, the number of each row ranges from 41 to 91. Piegl and Tiller's approach and Park and Wang's approach require 48 x 103 and 48 x 102 control points, shown in Fig. 11-(c), to interpolate all the rows, respectively.

**Table 5**
Comparison of the number of control points for a liver model by different number of rows

| m |  | 5 | 10 | 20 | 40 | 47 |
|---|---|---|---|---|---|---|
| min $n_i$ |  | 47 | 47 | 40 | 40 | 40 |
| max $n_i$ |  | 80 | 80 | 87 | 90 | 90 |
| Piegl's approach (q=5) | per=1.00 | 84 | 90 | 92 | 101 | 103 |
|  | per=0.75 | 98 | 132 | 145 | 172 | 173 |
|  | per=0.50 | 139 | 183 | 197 | 217 | 217 |
|  | per=0.25 | 218 | 294 | 332 | 403 | 412 |
| Park's approach (q=5) | per=1.00 | 85 | 87 | 92 | 101 | 102 |
|  | per=0.75 | 103 | 131 | 148 | 171 | 172 |
|  | per=0.50 | 144 | 178 | 193 | 211 | 212 |
|  | per=0.25 | 213 | 305 | 338 | 396 | 401 |

**Table 6**
Comparison of the number of control net for a human head with fixed per and different q

| m | min $n_i$ | max $n_i$ |  |
|---|---|---|---|
| 91 | 90 | 179 |  |
| per=1.0 | degree of row (q) | control net (Piegl) | control net (Park) |
|  | 2 | 92 x 182 | 92 x 184 |
|  | 3 | 92 x 183 | 92 x 182 |
|  | 4 | 92 x 184 | 92 x 186 |
|  | 5 | 92 x 185 | 92 x 184 |
|  | 6 | 92 x 186 | 92 x 188 |

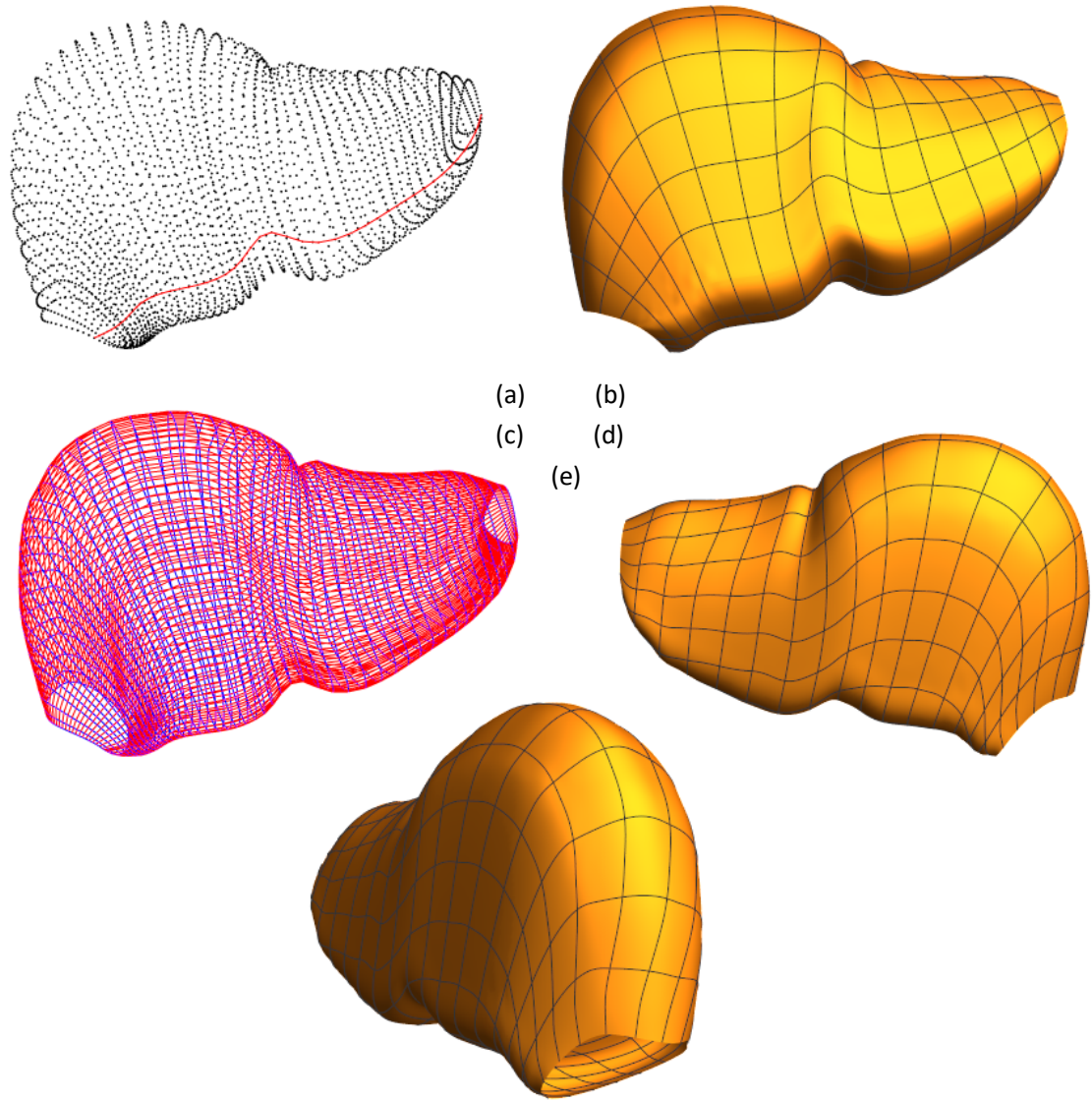

Fig. 11. Lofted B-spline surface to rows of data points sampled from sections of a liver model

(a) sampled points and based line; (b) lofted B-spline surface with *per*=1.0; (c) control net ; (d) – (e) surfaces at other angles.

Lastly, we give a comparison of Piegl and Tiller's [12] and Park and Wang's [13] algorithms. We use a fixed *per*=1.0 and let the degree of row ranges from 2 to 6, then applying them to the above two complicated models (human head and liver model). The corresponding results are summarized in Table 6 - 7, respectively.

**Table 7**

Comparison of the number of control net for liver-model with fixed *per* and different *q*

| $m$ | min $n_i$ | max $n_i$ | |
|---|---|---|---|
| 47 | 40 | 90 | |
| *per*=1.0 | degree of row($q$) | control net (Piegl) | control net(Park) |
| | 2 | 48 x 100 | 48 x 100 |
| | 3 | 48 x 102 | 48 x 102 |
| | 4 | 48 x 102 | 48 x 102 |
| | 5 | 48 x 104 | 48 x 104 |
| | 6 | 48 x 104 | 48 x 104 |

# 7. Conclusions and future work

This paper has presented some useful conjectures to address the issue that occurs in closed B-spline curve interpolation and lofted B-spline surface interpolation. We then apply these conjectures to the existing algorithms of lofted B-spline surface interpolation. The experimental results show the conjectures work well. Although the proposed conjectures have not been proven theoretically, our observation from numerical experiments, applications to the existing algorithms show that they are correct and feasible. The future work should be focused on two points, the proof of three conjectures proposed in this paper and the discovery of the necessary and sufficient conditions because the conjecture 1-2 just give sufficient conditions. We are also working on the proof of these three conjectures when completing this paper.